\documentstyle[prl,aps,epsfig]{revtex}
\begin{document}
\draft
\title{Nonextensive statistics, fluctuations and correlations in 
high energy nuclear collisions \footnote{Accepted for publication in Eur. Phys. J. C}}
\author{W. M. Alberico$^{1,3}$, A. Lavagno$^{2,3}$ and P. Quarati$^{2,4}$}
\address{
         $^1$Dipartimento di Fisica, Universit\`a di Torino, Via P. Giuria 1, 
             I-10124 Torino, Italy\\
         $^2$Dipartimento di Fisica, Politecnico di Torino, 
             C.so Duca degli Abruzzi 24, I-10129 Torino, Italy \\
         $^3$Istituto Nazionale di Fisica Nucleare, Sezione di Torino\\
         $^4$Istituto Nazionale di Fisica Nucleare, Sezione di Cagliari
        }

\maketitle

\begin{abstract}
Starting from the experimental evidence that high--energy nucleus--nucleus 
collisions cannot be described in terms of superpositions of elementary 
nucleon--nucleon interactions, we analyze the possibility that 
memory effects and long--range forces imply a nonextensive statistical 
regime during high energy heavy ion collisions. 
The relevance of these statistical effects and their compatibility with
the available experimental data are discussed.
In particular we show that theoretical estimates, 
obtained in the framework of the generalized nonextensive thermostatistics, 
can reproduce the shape of the pion transverse mass spectrum and explain 
the different physical origin of the transverse momentum 
correlation function of the pions emitted during the central Pb+Pb  
and during the p+p collisions at 158 A GeV. 
\end{abstract}
\vspace{0.5cm}

\section{Introduction}

In the last years, many efforts have been focussed on the study of high 
energy nuclear collisions, resulting in a better understanding of the strongly 
interacting matter at high energy density. Large acceptance detectors provide  
a detailed analysis of the multiplicity of produced particles; this in turn
can be related to possible signatures of the formation of a new phase of 
matter, the so--called quark--gluon plasma (QGP), in the early stages of 
high energy heavy ion collisions \cite{cse}.  

One of the most interesting experimental results is that 
high energy nucleus--nucleus (A+A) collisions cannot be described in terms 
of superpositions of elementary nucleon--nucleon (N+N) interactions 
(proton--proton or proton--antiproton). The different conditions of energy 
density and temperature at the early stage of the Pb+Pb and of the p+p 
collisions 
(to mention just two extreme examples) generate collective effects that 
modify the features of freeze--out observables. 
Of course the most appealing explanation would be to interpret 
the presence of these evident experimental differences as an indirect 
consequence of the formation of QGP in the early stages of heavy ion 
collisions.

It is an experimentally established fact \cite{bear} that the slope parameter 
of the transverse momentum distribution significantly increases with the 
particle mass when going from p+p to central Pb+Pb collisions. 
On the other hand, the preliminary results of the NA49 Collaboration 
\cite{rol,prep} indicate that the transverse momentum fluctuations in 
central Pb+Pb collisions at 158 A GeV have a quite different physical 
origin with respect to the corresponding fluctuations in p+p collisions. 
Let us also recall that the observed $J/\Psi$ 
suppression increases continuously and monotonically from 
the lightest (p+p) to the heaviest (S+U) interacting nuclei, but it 
exhibits a clear departure from this "normal" behavior for central 
Pb+Pb collisions \cite{NA50}.  

These experimental features have been explained in terms of collective 
effects in the hadronic medium 
(such as the presence of a transverse hydrodynamical expansion) 
and considering the strong influence of secondary rescatterings 
in heavy ion collisions \cite{bear,blei,blei2,heinz}. 
It is a very important issue to understand whether 
the system is able to reach full thermalization; 
up to now this question has been the subject of many experimental 
and theoretical studies. However several experimental observables, such 
as transverse mass spectrum, multiplicity of particles, three--dimensional  
phase--space density of pions, 
are well reproduced in the framework of local thermal equilibrium 
in a hydrodynamical expanding 
environment \cite{heinz,becca1,becca2,appel,E877}. 

In this paper we start from the hypothesis that memory effects and 
long--range forces can occur during high energy heavy ion collisions:
in this case  the strong influence of  collective effects on the 
experimental observables can be understood in a very natural way
in the framework of the generalized nonextensive thermostatistics. 
In this context we show that the  transverse momentum 
correlations of the pions emitted from central Pb+Pb collisions 
can be very well reproduced by means of very small deviations from the standard 
equilibrium extensive statistics. 

In Sec.\ref{kin_sec} we summarize the main  
assumptions contained in the derivation of the 
dynamical kinetical equations and their relevance to the 
determination of the equilibrium phase--space distribution. 
In Sec.\ref{qcd_sec} we examine whether these conditions are met in the 
high energy heavy ion collisions and which experimental observables 
can be sensitive to the presence of nonextensive statistical effects. 
In Sec.\ref{tsa_sec} we introduce the nonextensive Tsallis thermostatistics 
which is considered the natural generalization of the standard classical 
and quantum statistics when memory effects and/or long range 
forces are not negligible. 
In Sec.\ref{ptm_sec} we show how the generalized distribution function 
modifies the shape of the transverse mass spectrum and in Sec.\ref{phi_sec} 
we investigate the influence of  nonextensive particle fluctuations in the 
determination of the measure of the transverse momentum correlations. 
Finally, a discussion of the results and some conclusions are reported in 
Sec.\ref{conclu_sec}.

\section{Dynamical kinetical approach to thermal equilibrium and extensive statistics}
\label{kin_sec}

Assuming  local thermal equilibrium,  the freeze--out observables are
usually calculated within the extensive thermostatistics. In particular, 
the equilibrium transverse momentum distribution is assumed to be,  
in the classical case, the standard 
Maxwell--Boltzmann (J\"uttern) distribution. 
If the emitted  particles are light enough, such as pions, to have 
noticeable quantum degeneracy, one must use 
the Bose--Einstein (or Fermi--Dirac, for fermions) distribution.

When the system approaches  equilibrium, 
the phase--space distribution should be derived 
as a stationary  state of the dynamical kinetical evolution equation. 
If, for sake of simplicity, we limit our discussion to the classical 
case,  the Maxwellian distribution is obtained as a steady state solution of 
the Boltzmann equation.
Let us now briefly review the principal assumptions in the derivation of the 
Boltzmann equation and examine how these approximations 
can affect the determination 
of the equilibrium distribution \cite{bal}. 
One important assumption concerns the collision time, which must be  much 
smaller 
than the mean time between collisions. This request can be expressed by  
the condition $nr_0^3\ll 1$, where 
$n$ is the density and $r_0$ is the effective range of the interactions. 
This condition has two important physical consequences. 
a) There is no overlapping between subsequent collisions involving a given 
particle and the interactions can be described as a succession of simple 
binary collisions. b) It is always possible to define a time interval
in which the single particle distribution does not change appreciably and 
its rate of change at time $t$ depends 
only on its instantaneous value and not on its previous history. 
Hence this property reflects the Markovian character of the Boltzmann 
equation: no memory is taken into account. 

The second assumption is  analoguous to the first one, but for the 
space dependence of the  
distribution function: its rate of change  at a spatial point 
depends only on the neighborhood of that point. 
In other words the range of the interactions is short with respect to 
the characteristic spatial dimension of the system. 

The last important assumption is the so--called 
Bolzmann's Stosszahlansatz: the momenta of two particles 
at the same spatial point are not correlated and the corresponding 
two body correlation function  can be factorized 
as a product of two single particle distributions. 
This assumption is very important because it allows us 
to write down the collisional integral (which is equal to the total time 
rate of change of the distribution function)  in terms of the 
single particle distribution only. 
The above  assumptions, namely the  absence of non--Markovian memory 
effects, the absence of long--range interactions and negligible local 
correlations, together with  the Boltzmann's H theorem (based on the 
{\em extensive} definition of the entropy) lead us to the well--known 
stationary Maxwellian distribution. 

The basic assumption of  standard statistical mechanics is that the system 
under consideration can be subdivided into a set of non--overlapping 
subsystems. As a consequence the Boltzmann--Gibbs entropy is extensive 
in the sense that the total entropy of  two independent subsystems is the 
sum of their entropies. If memory effects and long range forces are present, 
this property is no longer valid and the entropy, which 
is a measure of the information about the particle distribution in the 
states available to the system, is not an extensive quantity. 

In the next section, we will see whether the above assumptions 
are implemented during high energy nuclear collisions and if there appear 
experimental signals that can be interpreted as a consequence of 
the presence of a nonextensive regime.

\section{Do heavy ion collisions satisfy extensive statistics?}
\label{qcd_sec}

It is a rather common opinion that, because of the extreme conditions of 
density and temperature in ultrarelativistic heavy ion collisions, 
memory effects and long--range color interactions give rise to the presence 
of non--Markovian processes in the kinetic equation affecting the 
thermalization process toward equilibrium as well as the standard 
equilibrium distribution \cite{gav,hei,biro,ropke}. 

A rigorous determination of the conditions that produce a 
nonextensive behavior, due to memory effects and/or
long--range interactions,  should be based on microscopic calculations 
relative 
to the parton plasma originated during the high energy collisions. 
At this stage we limit ourselves to consider the problem from a qualitative 
point of view on the basis of the existing theoretical calculations and 
experimental evidences.

Under the hypothesis that QGP is generated in the high energy collisions, the 
 quantities characterizing the plasma, such as lifetime and damping rates 
of  quasiparticles, are usually calculated within  finite temperature 
perturbative QCD. This approach is warranted by the fact that at high 
temperature (beyond a few hundred MeV) the strong QCD coupling 
 $\alpha_s= g^2/4\pi$ becomes very small and ``weak coupling'' regime 
takes place.  
In this regime ($g\ll 1$) the longitudinal gluon propagator is characterized 
by a Debye mass $m_{_D} \approx gT$, and the corresponding 
screening length ($\lambda_{_D}=m_{_D}^{-1}\approx 1/gT$) 
is much larger than the mean interparticle distance 
($\langle r \rangle \approx n^{-1/3}\approx 1/T$) 
(for a review, see e.g., Ref.\cite{bla}). Therefore a great parton number 
is contained in the Debye sphere and the ordinary mean field approximation 
of  QGP (Debye--H\"uckel theory) holds. 
Nevertheless, in the proximity of the phase transition, the QGP is no 
longer a system of weakly 
interacting particles and non--perturbative QCD calculations become important. 
Recently it has been shown \cite{schafer} that for temperature 
near the critical one ($T\ge 1.1 T_c$) non--perturbative calculations imply 
an effective quark mass very close to $T$, sensibly different from the value, 
proportional to $gT$, obtained in perturbative regime. Near the phase 
transition the characteristic quantity $nr_0^3\approx n\lambda_D^3$ 
should then be very close to one and only a small number of partons 
is present in the Debye sphere: the 
ordinary mean field approximation of the plasma is no longer correct 
and memory effects are not negligible. 
In addition, we observe that in high density quark matter the color magnetic 
field remains unscreened (in leading order) and long--range color magnetic 
interaction should be present at all temperatures.

From the above considerations it appears reasonable that, 
if the deconfining phase transition takes place, non--Markovian effects 
and long range interactions can influence the dynamical evolution of the 
generated fireball toward the freeze--out stage. Moreover they will 
affect the equilibrium phase--space distribution function if
thermal equilibrium is attained.
A signature of these effects should show up in physical observables. 
In this context, we notice that the authors of Ref.\cite{blei2} 
raise a controversy on   
the Markovian description of the multiple scattering processes, one of
the assumptions of the UrQMD model. This example  concerns
( $e^+ e^-\rightarrow \mu^+\mu^-$) scattering,  
where a determination of the angular distribution based on the 
Markovian approximation of the transport theory leads to wrong results 
\cite{bauer,bedia,ropke}. 

The aim  of the present work is to suggest a possible interpretation of
the above mentioned difference between the correlactions/fluctuations 
of the pion transverse momentum measured in p+p and Pb+Pb collisions 
\cite{appel} as a signature of the nonextensivity of the system.
In fact, it has been shown \cite{belka} that, whether the equilibrium 
condition is realized or not, the measure of the correlations/fluctuations 
strongly depends on  
the factorization of the multiparticle distributions and on the event
by event multiplicity of the  particles under consideration. Such property 
implies that the (multi)particle distribution functions are not inclusive 
(the distribution functions of one, two, or more bodies, depend on the 
presence of the other particles of the system). In turn, this is a 
manifestation of a nonextensive behavior of the system.

In addition we notice that, from a recent analysis of the average 
pion phase--space density at freeze--out in S--nucleus and  Pb--Pb 
collisions at SPS, the experimental data indicate a slower decrease 
with increasing $p_\perp$ than the Bose--Einstein curve \cite{ferenc}. 
The analysis of the same quantity in $\pi$--p collisions seems, instead, 
to be consistent with the standard expectations.
We remind that in Ref. \cite{ande} the authors show that 
by assuming the standard J\"uttern momentum distribution of emitted particles 
in high energy collisions one is led to unrealistic consequences and the 
physical freeze--out is actually not realized. 

From the foregoing considerations we conclude that both theoretical 
calculations and 
experimental observables agree with the existence of nonextensive 
features in high energy heavy ion collisions. 

\section{Generalized nonextensive Tsallis statistics}
\label{tsa_sec}

Several new developments in statistical mechanics have shown that in the 
presence of long--range forces and/or in irreversible processes related to 
microscopic long--time memory effects, the extensive thermodynamics, based 
on the conventional Boltzmann--Gibbs thermostatistics, is no longer correct 
and, consequently, the equilibrium particle distribution functions can show 
different shapes from the conventional well known distributions. 
Standard sums, or integrals, which appear in the calculation of 
thermostatistical quantities, like the partition function, the entropy, 
the internal energy, etc. can diverge. These difficulties are  
 well known in several physical domains \cite{lang} since long time.

A quite interesting generalization of the conventional Boltzmann--Gibbs 
statistics has been recently proposed by Tsallis \cite{tsa} and proves to 
be able to overcome the  shortcomings  of the conventional statistical 
mechanics in many physical problems, where the presence of 
long--range interactions, long--range microscopic memory, or fractal 
space--time constraints hinders the usual statistical assumptions. 
Among other applications, we quote astrophysical 
self--gravitating systems \cite{plasg}, the solar neutrino problem \cite{qua}, 
cosmology \cite{hami}, many--body, dynamical linear response theory 
and variational methods \cite{raja2}, phase shift analyses for 
the pion--nucleus scattering \cite{ion}.

The Tsallis generalized thermostatistics  is based upon the following 
generalization of the entropy \cite{tsa} 
\begin{equation}
S_q=\frac{1}{q-1}\, \sum_{i=1}^W p_i \, (1-p_i^{q-1}) \;,
\label{tsaen}
\end{equation}
where $p_k$ is the probability of a given microstate among $W$ different ones 
and $q$ is a fixed real parameter.

The new entropy has the usual properties of positivity, equiprobability, 
concavity and irreversibility, preserves the whole mathematical 
structure of thermodynamics (Legendre transformations) and
reduces to the conventional Boltzmann--Gibbs entropy $S=-\sum_i p_i \log p_i$
in the limit $q\rightarrow 1$.

The deformation parameter $q$ measures the  degree of  nonextensivity of the 
theory. In fact, if we have two independent systems $A$ e $B$,  such that
the probability of $A+B$ is factorized into 
$p_{A+B}(u_A, u_B)=p_A(u_A) \, p_B(u_B)$, the global entropy is not simply 
the sum of  their entropies  but it is easy to verify that 
\begin{equation}
S_q(A+B)=S_q(A)+S_q(B)+(1-q) S_q(A)S_q(B) \; .
\end{equation}

For a better appreciation of the meaning of nonextensivity in  a physical 
system, we discuss here another important property. Let us suppose that the 
set of $W$ microstates is arbitrarily separated into two subsets having
 $W_L$ and $W_M$ microstates ($W_L+W_M=W$), respectively, and 
define as $p_L \equiv \sum_{i=1}^{W_L}p_i$ and $p_M \equiv
\sum_{i=W_L+1}^{W}p_i$ the corresponding probabilities: hence $p_L+p_M=1$. 
It is then easy to show that
\begin{eqnarray}
S_q(\{p_i\})= &S_q(p_L,p_M)\; 
&+\;p_L^q\;S_q(\{p_i/p_L\})\;+\; p_M^q\;S_q(\{p_i/p_M\}) \; ,
\label{Shann}
\end{eqnarray}
where the sets $\{p_i/p_L\}$ and $\{p_i/p_M\}$ are the conditional
probabilities. This is a genera\-li\-zation of the famous Shannon's property 
but for the appearance of $p_L^q$ and $p_M^q$ instead of $p_L$ and $p_M$
in the second and third terms of the right hand side of (\ref{Shann}).
Since the probabilities $\{p_i\}$ are normalized, 
$p_i^q>p_i$ for $q<1$ and  $p_i^q<p_i$ for $q>1$: as a consequence values 
of  $q<1$ ( $q>1$ ) will favour rare ( frequent) events, respectively.

The single particle distribution function is obtained  through the 
usual procedure of maximizing the Tsallis entropy  under the 
constraints of keeping constant 
the average internal energy and the average number of particles. 
For a dilute gas of particles and/or for $q\approx 1$ values, the 
average occupational number can be written in a simple analytical 
form \cite{buyu} 
\begin{equation}
\langle n_i\rangle_q=\frac{1}{[1+(q-1)\beta (E_i-\mu)]^{1/(q-1)}\pm 1} \;,
\label{distri}
\end{equation}
where the  $+$ sign is for fermions, the $-$ for bosons and $\beta=1/T$.
In the limit 
$q\rightarrow 1$ (extensive statistics), one recovers the conventional 
Fermi--Dirac and Bose--Einstein distribution.

Let us  remind that the Tsallis generalized statistics does not entail 
a violation of the Pauli exclusion principle and does not modify the 
inclusive behavior of the bosons, but it modifies, with the 
generalized nonextensive entropy (\ref{tsaen}), the extensive nature of 
the standard statistics. At the equilibrium, the nonextensive statistics 
implies a finite temperature particle distribution different from the 
standard Fermi--Dirac and Bose--Einstein distributions; in the 
classical limit, one has the following  generalized Maxwell--Boltzmann 
distribution \cite{tsa}:
\begin{equation}
\langle n_i\rangle_q=[1+(q-1)\beta (E_i-\mu)]^{1/(1-q)} \;.
\label{distribz}
\end{equation}

When the entropic $q$ parameter is smaller than $1$, the distributions 
(\ref{distri}) and (\ref{distribz}) have a natural high energy cut--off: 
$E_i\le 1/[\beta (1-q)]+\mu$,  which implies that the energy tail is 
depleted; when $q$ is greater than $1$, the cut--off is absent and the energy 
tail of the particle distribution (for fermions and bosons) is enhanced. 
Hence the nonextensive statistics entails a sensible difference of the 
particle distribution shape in the high energy region with respect 
to the standard statistics.
This property plays an important r\^ole in the interpretation of the 
physical observables, as it will be shown in the following.

\section{Transverse mass spectrum and $q$--blue shift} 
\label{ptm_sec}

The applicability of the equilibrium statistical mechanics relies on
the recognition that the experimental transverse mass spectrum and 
the relative abundance of charged particles (pions, kaons, protons, etc.) 
can be described within the equilibrium formalism. 
The single particle spectrum can be expressed as an integral over a freeze--out 
hypersurface $\Sigma_f$ \cite{cooper}
\begin{equation}
E \frac{d^3N}{d^3p}=\frac{dN}{dy\, m_\perp dm_\perp d\phi}=
\frac{g}{(2\pi)^3} \int_{\Sigma_f} p^\mu d\sigma_\mu(x) f(x,p) \, ,
\end{equation}
where $g$ is the degeneracy factor and $f(x,p)$ is the phase--space distribution. 

Indeed, by assuming a purely thermal source with a Boltzmann distribution, 
the transverse mass spectrum can be expressed as follows:
\begin{equation}
\frac{dN}{m_\perp dm_\perp}=A \; m_\perp K_1\left (z \right ) \;, 
\end{equation}
where $z=m_{\perp}/T$ and $K_1$ is the first order modified Bessel function. 
In the asymptotic limit, $m_{\perp}\gg T$ ($z\gg 1$), the above expression 
gives rise to the  exponential shape 
\begin{equation}
\frac{dN}{m_\perp dm_\perp}=B \; \sqrt{m_\perp} \, e^{-z} \;, 
\label{mps}
\end{equation}
which is usually employed to fit the experimental 
transverse mass spectra and provides an indication of a thermal energy 
distribution. 

Although, from a general perspective, the experimental distribution of 
particle momenta in the transverse direction is described by an 
exponentially decreasing behavior, at higher energies 
the experimental data deviate from the usual Boltzmann slope. 
This discrepancy can be cured  by introducing the dynamical effect of 
a collective transverse flow 
which predicts the so--called blue shift factor, i.e., 
an increase of the slope parameter $T$ at large $m_\perp$.

The experimental slope parameter measures the particle energy, which 
contains both thermal (random) and collective (mainly due to rescattering) 
contributions. The thermal motion determines the freeze--out temperature 
$T_{fo}$, namely the temperature when particles cease to interact with 
each other. In the presence of a collective transverse flow \cite{bear}
one can extract $T_{fo}$ from the empirical relation
\begin{equation}
T=T_{fo}+ m \, \langle v_\perp\rangle^2 \; ,
\label{tslope}
\end{equation}
where $\langle v_\perp\rangle$ is a fit parameter which can be identified with the 
average collective flow velocity; the latter is consistent with zero 
in $p+p$ reactions, but can be large in collisions between heavy nuclei
(where rescattering is expected to be important).
In Eq.(\ref{tslope}), $m$ is the mass of the detected particle (e.g. pions, 
kaons, protons). 

Let us consider a different  point of view and argue that the deviation 
from the Boltzmann slope at high $p_\perp$ can be ascribed to the presence 
of nonextensive statistical effects in the steady state distribution of 
the particle gas; the latter do not exclude the physical effect of a 
collective flow but rather incorporates a description of it from a 
slightly different statistical mechanics analysis.

If long tail time memory and long--range interactions are present, as already 
discussed in Sec.\ref{qcd_sec}, 
the Maxwell--Boltzmann distribution must be replaced by
the generalized distribution (\ref{distribz}).
We consider here only small deviations from 
standard statistics ($q-1\approx 0$); then 
at  first order in 
$(q-1)$ the transverse mass spectrum can be written as 
\begin{equation}
\frac{dN}{m_\perp dm_\perp}=C \; m_\perp \left\{ K_1\left ( z \right )+ 
\frac{(q-1)}{8} z^2 \; \left [3 \; K_1 (z)+K_3 (z) \right ]  \right\} \;,
\label{mpt}
\end{equation}
where $K_3$ is the modified Bessel function of the third order. 
In the asymptotic limit, 
%$m_{\perp}\gg T$, 
$z\gg 1$ (but always for $q$ values close to 1, such that $(q-1)z\ll 1$), 
we obtain the following generalization of Eq.(\ref{mps}):
\begin{equation}
\frac{dN}{m_\perp dm_\perp}=D \; \sqrt{m_\perp} \, 
\exp \left (-z+\frac{q-1}{2}\, z^2\right )\;. 
\label{mpst}
\end{equation}

From the  above equation it is easy to see that at  first 
order in $(q-1)$ the generalized slope parameter becomes the quantity $T_q$, 
defined as 
\begin{equation}
T_q=T+ (q-1) \, m_\perp \; .
\label{qtslope}
\end{equation}
Hence nonextensive statistics predicts, {\it in a purely thermal source}, 
 a generalized $q$--blue shift factor 
at high $m_\perp$; moreover  this shift factor is 
not constant but increases (if $q>1$) with $m_\perp=\sqrt{m^2+p_\perp^2}$, 
where $m$ is the mass of the detected particle. We have in this case 
an effect very similar to the one described by the phenomenological 
Eq.(\ref{tslope}). 
We remark that the proposed distribution of Eq.(\ref{mpst}), 
based on the nonextensive 
thermostatistics, has the same high $p_\perp$ dependence of 
the parameterization 
reported in Ref.\cite{appe} to fit the transverse distribution for central 
Pb+Pb collisions. 

The value of the entropic parameter $q$ is a measure of the 
nonextensivity of the system and, as a consequence, it should be fixed by the 
reaction under consideration. On the basis of the above  considerations, 
at a fixed reaction energy, one should find that larger value of $q$ 
are required for heavier 
colliding nuclei, because at higher energy density we have a 
larger probability for the phase transition to be  achieved.

A quantitative description of the particle production in nonextensive 
statistics goes beyond of the scope of this paper;  it is fair 
to assume that Eq.(\ref{mpst}), for $q>1$ value, can be in good 
agreement with experimental results in heavy ion collisions. 
 In this framework, indeed,  one can understand the 
slower decrease (for heavier colliding nuclei) of the pion phase--space 
distribution  with increasing $p_\perp$.

It is worth noticing  that for small deviations from
 the standard statistics ($|(q-1)|<0.05$) the experimental midrapidity 
transverse mass distributions are well reproduced with the same slope 
parameter of Ref. \cite{bear}, within the statistical errors. 

In the next Section we shall calculate the measure of the transverse 
momentum fluctuations in the framework of the equilibrium nonextensive 
statistics. We notice that, since small nonextensive statistical effects 
are consistent with the experimental transverse mass spectrum, 
 we can use, in what follows, the same temperature extracted 
from the standard analysis of the experimental results.

\section{Transverse momentum fluctuations in nonextensive statistics}
\label{phi_sec}

In Ref.\cite{gaz92}, Ga\'zdzicki and Mr\'owczy\'nski have introduced a  
quantity $\Phi$, which  measures the event--by--event fluctuations and does 
not explicitly depend on the particle multiplicity, but 
 appears to be sensitive to the correlations.
It is defined as  
\begin{equation}
\Phi_x=\sqrt{\langle Z_x^2 \rangle \over \langle N \rangle} -
\sqrt{\overline{z_x^2}} \;,
\label{phix}
\end{equation}
where $x$ is a variable such as energy or transverse momentum, 
$z_x = x - \overline{x}\,$ is a single particle variable and 
$Z_x = \sum_{i=1}^{N}(x_i - \overline{x})$ is the corresponding 
sum--variable,  $N$ being the number of the particles in the event.

A non--vanishing value of the measure $\Phi$ implies  that 
there is an effective correlation among particles (dynamical or statistical) 
which alters the momentum distribution.
%depends on the particle multiplicity. 
It is usually assumed that if the experimental value of $\Phi$ turns out 
to be constant in going from  N+N to A+A collisions, then 
nucleus--nucleus  collisions  can be described as an 
incoherent superposition of elementary N+N scattering processes 
and the correlations among particles are unchanged \cite{belka,gaz92}.
We wish here to point out that this might not be the case: indeed it will
be shown that similar values of $\Phi$ can be attributed to different 
statistical correlations, which in turn reflect distinct dynamical
situations.

Recently the correlation measure $\Phi_{p_\perp}$ of the pion transverse 
momentum has been measured by NA49 Collaboration. The experimental results 
correspond to a value $\Phi^{exp}_{p_\perp}=0.6\pm 1$ MeV for Pb+Pb at 
158 A GeV and to 
$\Phi^{exp}_{p_\perp}=5\pm 1$ MeV in the preliminary measurements 
of the p+p collisions at the same energy \cite{rol,prep}. However,
in their most recent analysis of the Pb+Pb data, the NA49 Collaboration 
has estimated a contribution of $\Delta\Phi_{p_\perp}=5\pm 1.5$~MeV 
from the statistical two--particle correlation function alone and 
an ``anti--correlation'' contribution of $\Delta\Phi_{p_\perp}=-4\pm 0.5$~MeV, 
stemming from the limitation in the two--track resolution of the NA49 
apparatus \cite{prep}. 
Hence it appears that the actual physical values both for p+p 
and for Pb+Pb collisions are nearly equal. 
In any case, as it was observed in the same 
%experimental 
paper, the physical origin of transverse momentum fluctuation 
remarkably changes from Pb+Pb to p+p collisions and, although 
the pion transverse correlation measure has been studied within many 
different theoretical approaches, the obtained results are somewhat 
controversial and not well understood \cite{blei,mro,gaz97,gaz98,capella}. 

We will show in the following that the physical differences in the 
origin  of the pion transverse momentum fluctuations in Pb+Pb collisions 
with respect to the p+p ones can be understood in the framework of the 
nonextensive statistics. For this purpose, keeping in mind that
the whole mathematical structure of the thermodynamical relations is
 preserved in the nonextensive statistics, it is easy to show that the 
two terms in the right hand side of Eq.(\ref{phix}) can be expressed in 
the following simple form
\begin{eqnarray}
\overline{z^2_{p_{\perp}}} =
{1 \over \rho}\int{d^3p \over (2\pi )^3} \, \, \Big(p_{\perp} - 
\overline{p}_{\perp} \Big)^2   \langle n \rangle_q \; ,
\label{qz2p}
\end{eqnarray}
and 
\begin{eqnarray}
{\langle Z_{p_{\perp}}^2 \rangle \over \langle N \rangle }=
{1 \over \rho}\int{d^3p \over (2\pi )^3}
\,\Big(p_{\perp} - \overline{p}_{\perp} \Big)^2   \langle\Delta n^2\rangle_q \;,
\label{qz2g}
\end{eqnarray}
where the quantity $\overline{p}_{\perp}$ is the average transverse momentum,
\begin{equation}
\overline{p}_{\perp} = {1 \over \rho}\int{d^3p \over (2\pi )^3} \;
p_{\perp} \langle n \rangle_q \ \ \ \ \ \ \  {\rm with} 
\ \ \ \ \ \ \ \rho=\int{d^3p \over (2\pi )^3} \langle n \rangle_q \;.
\end{equation}
In the above equations we have indicated with $\langle n \rangle_q$ the mean 
occupation number of Eq.(\ref{distri}) extended to the continuum and with 
$\langle\Delta n^2\rangle_q=\langle n^2\rangle_q-\langle n \rangle^2_q$ the 
generalized particle fluctuations, given by 
\begin{equation}
\langle\Delta n^2\rangle_q\equiv
\frac{1}{\beta}\frac{\partial\langle n\rangle_q}{\partial\mu}=
\frac{\langle n\rangle_q }{1+(q-1)\beta (E-\mu)}\, (1\mp \langle n\rangle_q)\;,
\label{fluc}
\end{equation}
where $E$ is the relativistic energy $E=\sqrt{m^2 + p^2}$. 
For $q=1$, one recovers the well--known text--book  expression for the 
fluctuations of fermions ($-$) and bosons ($+$). 
 Eq.(\ref{qz2g}) shows that the correlation measure $\Phi$ arises from 
the particle fluctuations; this result does not come as a surprise 
since  it is well known that the particle fluctuations are related to the 
two--body density or to the isothermal compressibility by means of the 
so--called Ornstein--Zernike relation \cite{bal}. 

In the standard extensive statistics ($q=1$), one obtains that the 
correlation measure $\Phi$ is always negative for fermions and positive 
for bosons, while it vanishes in the classical standard statistics \cite{mro}. 
Indeed  $\langle Z^2\rangle$ is defined 
in terms of the fluctuations $\overline{\Delta n^2}$, which, for the ideal 
Fermi--Dirac (Bose--Einstein) gas, are suppressed (enhanced) with respect 
to the classical case. 
In the nonextensive Tsallis statistics, the fluctuations 
of an ideal gas of fermions (bosons), expressed 
by Eq.(\ref{fluc}), are still suppressed (enhanced) by the factor 
$1\mp\langle n\rangle_q$; this effect, however, is modulated by the 
factor $[1+(q-1) \beta (E-\mu)]^{-1}$ in the right hand side 
of Eq.(\ref{fluc}). Therefore the fluctuations turn out to be increased 
for $q<1$ (we remind that due to the energy cut--off condition 
the above factor is always positive) 
and are decreased for $q>1$. The measure $\Phi$ 
does not have, both for bosons and fermions, 
a constant sign and it is positive or negative 
depending on the value of the entropic parameter $q$. 
In the classical nonextensive case  
we also have  $\langle\Delta n^2\rangle_q\ne \langle n\rangle_q$. 
The quantity $\langle\Delta n^2\rangle_q$ 
is greater than $\langle n\rangle_q$ for $q<1$, and smaller for 
$q>1$ and the correlation measure $\Phi$  
can be different from zero for $q\ne 1$. 
This is essentially due to the intrinsic nature of the system in 
the presence of memory effects and/or long--range forces  
and to the dependence of the (multi)particle distribution functions 
upon the surrounding medium.  

We use Eqs.(\ref{qz2p}) and (\ref{qz2g}) 
to evaluate  the correlation measure $\Phi_{p_\perp}$ for the 
pion gas ($m_\pi=140$ MeV), which we can compare with the 
experimental results measured in the central 
Pb+Pb collision by the NA49 collaboration \cite{rol,prep}. 
Let us notice that the values of $\Phi_{p_\perp}$, which 
depends on  $T$, $\mu$ and the entropic parameter $q$, are very sensitive 
to  small variations of $q$. 
This is mainly due to the modified expression (\ref{fluc}) of the 
fluctuations in the nonextensive statistics. 

\begin{figure}[htb]
\mbox{\epsfig{file=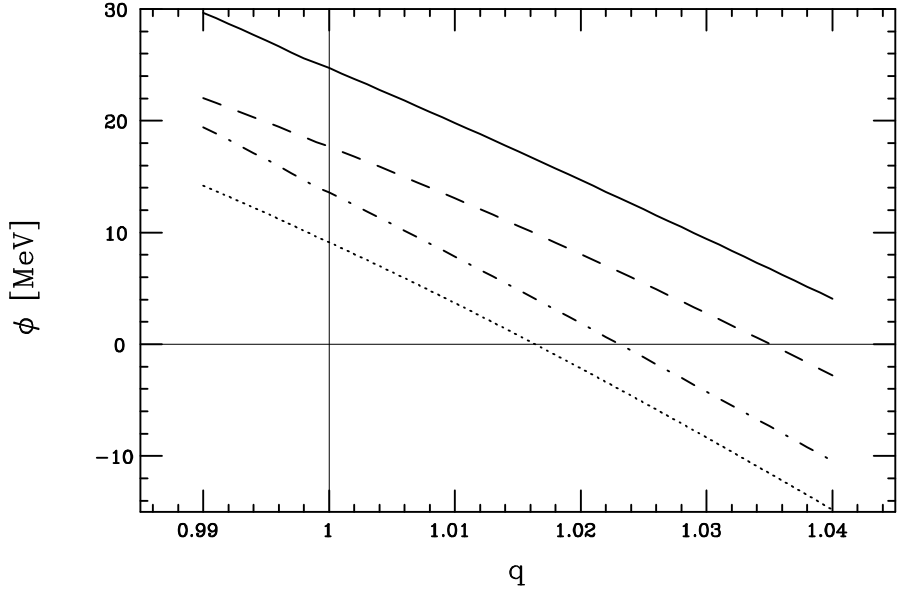,width=0.8\textwidth}}
%\vspace{0.2cm}
\caption[]{The correlation measure $\Phi_{p_\perp}$ [MeV] as a function of 
the entropic  parameter $q$, calculated in four  different conditions 
 of  temperature $T$ and chemical potential 
$\mu$ ($T=170$ MeV and $\mu=60$ MeV solid line, 
$T=140$ MeV and $\mu=60$ MeV dashed line, 
$T=170$ MeV and $\mu=0$ MeV dot-dashed line, 
$T=140$ MeV and $\mu=0$ MeV dotted line). 
The value $\Phi_{p_\perp}(q=1)$ corresponds to the correlation measure 
calculated within the standard extensive statistics.
}

\vspace{0.1cm}
\label{qmute}
\end{figure}

In Fig. \ref{qmute} we show $\Phi_{p_\perp}$ as a function of the 
parameter $q$, at different (fixed) values of the temperature and 
chemical potential.  
The experimental data are reported up to a maximum  value of the 
transverse momentum $p_\perp = 1.5$~GeV: hence 
we have limited the integration up to the upper limit of $1.5$ GeV. 
Concerning the temperature $T$, we use the slope parameter 
derived from the fit of the 
transverse momentum spectra \cite{bear,appe}. 
With $T=170$~MeV and   $\mu=60$ MeV  (according to the suggestion of 
\cite{shuflu}),  we obtain 
$\Phi_{p_\perp}= 5$ MeV with $q=1.038$, 
while the usual extensive statistics, with $q=1$, gives 
$\Phi_{p_\perp}=24.7$ MeV (we remind that a non--vanishing value of $\mu$
implies that there is no chemical equilibrium).

By choosing $\mu=0$ MeV one can obtain the same value $\Phi_{p_\perp}=5$ 
MeV with a smaller $q$ ($q=1.015$). In this case, with $q=1$, 
$\Phi_{p_\perp}=13.6$ MeV.

Thus, a very little deviation from the standard Bose--Einstein 
distribution is sufficient to obtain theoretical estimates in
agreement with the experimental results. 
We notice also that the value employed, $q\approx 1$, justifies, a 
posteriori, the approximation of the analytical form of the distribution 
function (\ref{distri}) and the validity of Eq.(\ref{mpt}) 
for the transverse mass spectrum. 
The correlations are much more affected by deviations from the extensive 
statistics than the single particle observables, such as the transverse 
mass distribution. We can obtain a very good agreement with the 
experimental results of $\Phi_{p_\perp}$, still being consistent
 with the other single particle measurements. 

\begin{figure}[htb]
\mbox{\epsfig{file=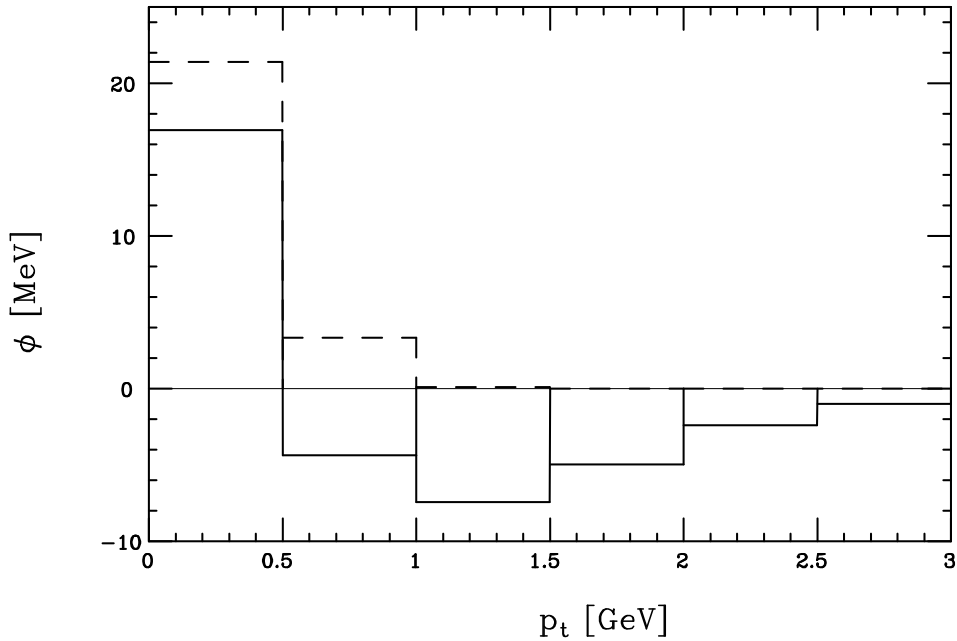,width=0.85\textwidth}}
%\vspace{0.2cm}
\caption[]{The partial contributions to the correlation measure 
$\Phi_{p_\perp}$ [MeV] in
 different $p_\perp$ intervals, at $T=170$ MeV and $\mu=60$ MeV. The dashed 
line refers to standard statistical calculations with $q=1$, the
solid line corresponds to 
$q=1.038$ (note that by summing the contributions up to 
$p_\perp=1.5$~GeV we obtain  the
 value  $\Phi_{p_\perp}=5$ MeV,  indicated by 
the NA49 experiment\cite{prep}. Negative contributions, at 
high $p_\perp$ steps, are predicted only within the nonextensive statistics.}
\label{isto}
\end{figure}

In this context, it is worth noticing that, even for small deviations from 
the standard statistics, an appreciable difference in the distribution 
function and in the fluctuations (with
respect to the standard ones) appears {\it especially} in the region of 
high transverse momenta. 
This behavior is consistent with the interpretation of 
the nonextensive statistics effects, since 
 the detected high  energy pions should be essentially the
primary ones, hence more sensitive to the early stage of the collisions. 
In order to show this fact,
 we have evaluated the partial contributions to the 
quantity $\Phi_{p_\perp}$, by using Eq.s (\ref{qz2p}) and (\ref{qz2g}), 
and by extending the integration over $p_\perp$ to partial
intervals $\Delta p_\perp=0.5$~GeV. The results are shown  in Fig.\ref{isto}, 
at $T=170$ MeV and $\mu=60$ MeV.   
In the standard statistics (dashed line), 
$\Phi_{p_\perp}$ is always positive and vanishes in the $p_\perp$--intervals 
above $\approx 1$ GeV. In the nonextensive statistics (solid line), instead, 
the fluctuation measure  $\Phi_{p_\perp}$ 
becomes negative for $p_\perp$ larger than $0.5$ GeV 
and becomes vanishingly small only 
 in $p_\perp$--intervals above $\sim 3$ GeV.

In view of these considerations, we suggest an 
analysis of the Pb+Pb measurements which allows to disentangle the 
contributions  to $\Phi_{p_\perp}$ in different $p_\perp$'s intervals. 
A negative value of $\Phi_{p_\perp}$ at high transverse momentum intervals 
could be a clear evidence of the presence of a nonextensive regime.

\section{Conclusions}
\label{conclu_sec}

The nonextensive statistics appears suitable to evaluate physical observables
 recently measured  in heavy ion collision experiments. 
The spectrum of the transverse momentum can be reproduced by means of 
the  nonextensive distribution, which naturally takes into account  typical 
collective effects in heavy colliding nuclei such as the increasing of the 
slope parameter with high $p_\perp$ and  high particle masses.
The calculated correlation measure $\Phi_{p_\perp}$ agrees with the 
experimental value by considering 
 a  small deviation from the standard statistics.
Within the nonextensive approach we have also found 
negative partial contributions to $\Phi_{p_\perp}$ at high $p_\perp$ 
(larger than $0.5$ GeV). In this regime 
the $\Phi_{p_\perp}$ value is not affected by resonance decays: 
hence an experimental confirmation of our prediction
would be an unambiguous  signal of the validity of nonextensive 
statistics in relativistic heavy ion collisions. 

The quantity $q$ is not just a free parameter: it depends on the physical 
conditions generated in the reaction and, 
in principle, should be related to microscopic 
quantities (such as the mean interparticle interaction length, the 
screening length and the frequency collision into the parton plasma). 
We have found that a value of $q$ slightly larger than one characterizes  
the correlation measure $\Phi$, both by suppressing
the boson fluctuations in Eq.(\ref{qz2g}) and  by enhancing the importance  
of the region of large transverse momenta in Eq.(\ref{qz2p}). 
In the diffusional approximation, a value $q>1$ implies the presence of 
a superdiffusion among the constituent particles 
(the mean square displacement is not linearly 
proportional to time, having, instead, a power law behavior 
$\langle x^2\rangle\propto t^\alpha$, with $\alpha>1$).  
Effects of anomalous diffusion are 
strongly related to the presence of non--Markovian memory interactions 
and colored noise force in the Langevin equation  \cite{tsamem}. In this 
context, we note that in Ref.\cite{jan}, the QCD chiral phase transition is 
analyzed in analogy with the metal--insulator one and it is shown 
that an anomalous diffusion regime can take place. 
The meaning of anomalous diffusion can be found in the presence of 
multiparticle rescattering, which  is very large 
in Pb+Pb collisions \cite{blei,blei2}. 

The early stage of the p+p collision is believed to be radically different 
from  the Pb+Pb case and the influence of the nonextensive statistics 
appears not negligible in the last reaction. 
In these terms the anomalous 
medium  effects and the remarkable experimental differences 
between light and heavy ion collisions should be better understood. 

Finally we notice that, for a given
value of the deformation parameter $q$, the
greater is the mass of the detected particle,  the
more strongly modified are the fluctuations.
For this reason, the present calculation predicts more important 
nonextensive effects for the correlations of heavier mesons and baryons, 
than for light particles: 
 future measurements of $\Phi_{p_\perp}$ for 
these heavier particles should give a more clear evidence of this effect.
Furthermore, the nonextensive statistics could imply a sizable
modification of the hadronic mass compressibility, 
since this quantity is strongly
related to particle fluctuations (being 
more sizably suppressed for particles of heavier masses\footnote{Using
the same value, $q=1.02$, of the deformation parameter, we can estimate a 
reduction of the standard Bose--enhancement factor $1+\langle n\rangle$ of 
about $2\%$ for pions, of $7\%$ for kaons and of $11\%$ 
for $\rho$ mesons.}) \cite{mro2}. 
We also remind that the nonextensive statistics implies, for $q>1$, 
a suppression of the standard boson fluctuations given in Eq.(\ref{fluc}),
which is more effective at high transverse momentum whereas it 
appears negligible for low momenta.  
Shuryak suggests\cite{shuflu} that  enhanced fluctuations for low 
$p_\perp$ bins, in the pion $p_\perp$ histograms, are a possible manifestation 
of disoriented chiral condensation. We may argue that 
this effect could be much more evident 
if one assumes the validity of the generalized statistics. 
New data and investigations are necessary to gain a deeper 
understanding of the high energy heavy--ion observables.

\vspace{0.5cm}

\noindent
{\bf Acknowledgements}\\
We are indebted to St. Mr\'owczy\'nski for useful suggestions, advices and 
comments to the original manuscript; we thank F. Becattini, 
D. Bresolin, G. Gervino, G. Roland and 
C. Tsallis for fruitful comments and discussions.

\end{document}